\documentclass[aps,prl,twocolumn,groupedaddress,floatfix,showpacs]{revtex4}
\usepackage{graphicx}
\usepackage{amsmath}

\begin{document}

\title{Pumping spin with electrical fields}

\author{M. Governale$^1$, F. Taddei$^{2,3}$, and Rosario Fazio$^3$}

\affiliation{$^1$Institut f\"ur Theoretische Festk\"orperphysik,
Universit\"at Karlsruhe, D-76128 Karlsruhe, Germany\\
$^2$ ISI Foundation, Viale Settimio Severo, 65, I-10133 Torino, Italy\\
$^3$ NEST-INFM \& Scuola Normale Superiore, I-56126 Pisa, Italy}

\date{\today}
\begin{abstract}
Spin currents can be obtained through adiabatic
pumping by means of electrical gating only.
This is possible by
making use of the tunability of the Rashba spin-orbit coupling
in semiconductor heterostructures.
We demonstrate the principles of this effect by considering a single-channel 
wire with a constriction. We also consider realistic structures, consisting 
of several open channels where subband spin-mixing and disorder are present,
and we confirm our predictions.
Two different ways to detect the spin-pumping effect, either using
ferromagnetic leads or applying a magnetic field, are discussed.
\end{abstract}


\maketitle
The investigation of spin-dependent transport  and its application
to steer electrical currents is at the foundation of 
\emph{Spintronics}~\cite{wolf}. Both of fundamental interest and 
of practical importance, the success in operating spin-based devices
relies on the ability to produce and control spin
currents.
At present, techniques to obtain a spin current 
include injection from ferromagnets~\cite{dassarma}, Zeeman's splitting 
due to magnetic fields, and 
optical excitations~\cite{ganichev}. 
Very recently some alternative proposals have been put forward.
Mucciolo {\em et al.}~\cite{mucciolo} suggested
to obtain spin currents based on the use of pumping of electrons 
through a chaotic dot in the presence of an in-plane magnetic field;
Brataas {\em et al.}~\cite{brataas} proposed a spin battery relying 
on a ferromagnet with precessing magnetization.

Adiabatic charge pumping~\cite{brouwer,zhou,switkes} 
consists in the transport of charge obtained, at zero 
bias voltage, through the periodic modulation of some parameters
(e. g. gate voltages) in the scattering region. 
If the time variation of the scattering 
matrix occurs on a long time scale compared to the transport time   
then the charge transferred per period does not depend on the detailed 
time evolution of the scattering matrix  but only on geometrical 
properties of the pumping cycle~\cite{brouwer}. 
Numerous works ( e.g. Refs~\cite{makhlin,levitov,levinson,aleiner,moskalets,butti} 
and references therein) addressed different aspects of
adiabatic pumping as, for example, the counting statistics of the 
pumped current, the generalization to multi-terminal geometries and 
the question of the phase coherence. 

Adiabatic pumping of spin seems to be quite attractive as well,
although little attention has been payed to it so far (see however 
Ref.\cite{mucciolo}). A combined implementation of
adiabatic charge pumping with a spin filter will ensure 
that if charge transport occurs also spin is transferred. 
In this Letter we discuss the possibility of spin pumping
without using
ferromagnetic materials or external magnetic fields. 
This is indeed possible by making use of the tunability 
of the Rashba effect~\cite{rashba,lommer,note0}.
A {\em spin current} is then produced by {\em electrical gating} only. 
Adiabatic pumping plays a crucial role in the present
mechanism since there is 
no spin-polarized current if the same device is dc biased and no 
time-dependent transport is involved.  
 
\begin{figure}
\includegraphics[width=3.in]{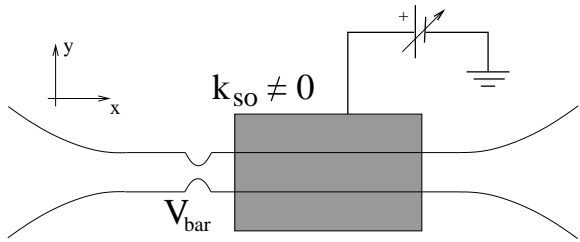}
\caption{Schematic setup for the adiabatic spin pump. It consists of 
a quantum wire with Rashba spin-orbit coupling $k_{\text{so}}$ modulated by a
gate (gray region) and controlled through a time-dependent
voltage generator.
A potential barrier $V_{\text{bar}}$ (represented by the constriction) is 
present at the interface between the left lead and the wire.}
\label{setup}
\vspace{1cm}
\end{figure}

Electrons confined in a two-dimensional electron gas, realized in 
a semiconductor heterostructure with some asymmetry in growth 
direction $z$, are subject to the Rashba spin-orbit (SO) 
coupling whose Hamiltonian reads 
$
H_{\text{so}} = \frac{\hbar k_{\text{so}}}{m}\left( \sigma_x\,
p_y - \sigma_y\, p_x\right),   
$
where 
$m$ is the effective mass. 
It is important to notice that the strength of the SO coupling, 
denoted as $k_{\text{so}}$,  
can be tuned by changing the asymmetry of the quantum well via 
externally applied voltages, as shown in several experimental 
studies~\cite{nitta,schaepers,grundler}. 
The system we have in mind to produce a spin current is 
schematically depicted in 
Fig.~\ref{setup}. It consists of a 
quantum wire (parallel to the $x$-axis) of length $L$ 
with Rashba spin-orbit coupling, connected to two semi-infinite 
leads, where spin-orbit coupling is absent. 
At the interface between the wire and the left lead a constriction 
will give rise to a potential barrier denoted by $V_{\text{bar}}$.
The Hamiltonian of the wire can be written as $H=H_{\text{1D}}+
H_{\text{tras}}+
H_{\text{mix}}$, with
\begin{subequations} 
\begin{eqnarray}
\label{h1d}
H_{\text{1D}}&=&\frac{1}{2m} p_x^2 -\frac{\hbar k_{\text{so}}}{m}
\sigma_y\, p_x,\\
\label{htras}
H_{\text{tras}}&=&\frac{1}{2m} p_y^2+ V_{\text{conf}}(y)\\
\label{hmix}
H_{\text{mix}}&=& \frac{\hbar k_{\text{so}}}{m}
\sigma_x\, p_y,
\end{eqnarray}
\end{subequations}
where $H_{\text{1D}}$ describes the longitudinal motion along the wire, 
$H_{\text{tras}}$ is the transverse part of the Hamiltonian (it contains 
the transverse confining potential $V_{\text{conf}}$), and 
$H_{\text{mix}}$ the part of the SO coupling 
that is responsible for subband mixing \cite{mireles, rashbawire}.
If the spin precession length $l_{\text{so}}=\pi/k_{\text{so}}$ is 
much larger than the typical width of the wire then $H_{\text{mix}}$ 
can be neglected and a 
common spin-quantization axis can be found (perpendicular to the 
wire and the heterostructure growth direction). In this limit 
the quasi-one-dimensional subband dispersion relations read:
$
\epsilon_{n,\sigma}=\frac{\hbar^2}{2 m}(k_x-\sigma k_{\text{so}})^2-
\Delta_{\text{so}}+E_n, 
$
where $E_n$ is the transverse energy, $\sigma=\pm$ is the quantum number 
of $\sigma_y$, 
and  $\Delta_{\text{so}}=\hbar^2 k_{\text{so}}^2/ 2m$.  
In order to compute the pumped charge and spin the 
scattering matrix should be determined. 
For the sake of simplicity and 
clarity we present the general idea considering that only a single subband  
is occupied both in the wire and leads.  
To avoid cluttering the notation and allow for simple analytical 
expressions, we further assume that no Fermi-velocity mismatch is present 
between the leads and the wire and that 
$\Delta_{\text{so}}$ is much smaller then the Fermi energy \cite{note1}.  
The analytical results are presented for a delta function potential 
$V_{\text{bar}}(x)=V \delta(x)$.
All these assumptions do not affect the basic principles of our proposal.
Indeed we will show that when these hypothesis are
relaxed, only small quantitative changes occur.
Since in this idealized model there is no spin-mixing mechanism, 
we can treat the two spin species separately. 
The pumping cycle is obtained by varying in time the height of 
the barrier (we define 
$\bar{V}=2m V /\hbar^2 $), 
and the spin-orbit coupling $k_{\text{so}}$. 
In the following we discuss both the average current and the noise spectrum.
The average  spin-$\sigma$ particle current is 
given by~\cite{brouwer}
\begin{equation}
\label{isigma}
I_{\sigma}
=\frac{\omega}{2 \pi^2}\int d\bar{V}\,dk_{\text{so}} 
D_\sigma(\bar{V},k_{\text{so}}), 
\end{equation}
where the integral is over the surface spanned during a cycle in parameter 
space, $\omega$ is the frequency of the pumping fields, and  
\begin{equation}
\label{dfactor}
D_\sigma(\bar{V},k_{\text{so}})=\text{Im}\left\{ 
\frac{\partial r_{\sigma}^{\prime *}}{\partial{\bar{V}}}
\frac{\partial r'_{\sigma}}{\partial{k_{\text{so}}}}+
\frac{\partial t_{\sigma}^{*}}{\partial{\bar{V}}}
\frac{\partial t_{\sigma}}{\partial{k_{\text{so}}}}\right\}
\end{equation}
with $r'_\sigma$ and $t_\sigma$ being, respectively, 
the reflection and transmission coefficient for an electron with 
spin $\sigma$. 
From the spin-$\sigma$ currents Eq.~(\ref{isigma}), we can define 
a charge$\backslash$spin current as 
$I_{\text{charge}\backslash\text{spin}}=I_+ \pm I_-$
(note that $I_{\text{charge}}$ is expressed in units of electron charge).

While the average pumped currents depend only on the geometrical properties 
of the pumping cycle, the current noise depends on the full 
time-dependence of the pumping parameters. 
Since we are interested in current fluctuations around the average current
we calculate the zero-frequency component of the noise spectrum 
\begin{equation}
\label{noise}
S_{\sigma}=\frac{\omega}{2 \pi}\int d\tau  \int_0^{\frac{2 \pi}{\omega}} 
d\tau^\prime 
\langle \Delta \hat{I}_\sigma (\tau) \Delta \hat{I}_\sigma (\tau^\prime)
\rangle,  
\end{equation}
where $\Delta \hat{I}_\sigma= \hat{I}_\sigma -\langle I_\sigma\rangle$. 
In the case of weak pumping  
the knowledge of the  average number of transmitted particles and 
of the zero-frequency noise characterizes the full counting 
statistics~\cite{levitov}. 
As there are no correlations between electrons with different spin indexes, 
the noise of the charge current and of the spin current is simply 
$S_{\text{spin}}=S_{\text{charge}}=S_+ + S_-$. 
Several authors have studied  noise in quantum pumps 
\cite{makhlin,moskalets,levitov}. We make use of the formulation of 
Moskalets {\it et al.}.

Once the scattering matrix is determined,
Eq.~(\ref{dfactor}) yields
\begin{equation}
\label{dsigma}
D_{\sigma}(\bar{V},k_{\text{so}})=\sigma \frac{4k_{\text{F}}^2 L \bar{V}}
{(4 k_{\text{F}}^2+\bar{V}^2)^2}, 
\end{equation} 
where $k_{\text{F}}$ is the Fermi wave-vector in the leads \cite{note2}. 
From Eq.(\ref{dsigma}) we immediately obtain that the  pumped 
charge current is zero and the pumped spin current is $I_{\text{spin}}=2I_+$.
For a sinusoidal pumping cycle:
$
\bar{V}=V_0+\Delta V \sin
(\omega \tau)\quad 
$
and
$ 
k_{\text{so}} =  k_{\text{so},0}+\Delta k_{\text{so}}
\sin (\omega \tau-\phi) 
$,
with $\Delta V \ll V_0$ (weak-pumping limit) we can determine the 
explicit form of the average current
\begin{equation}
I_{\text{spin}}=
\frac{\omega}{2 \pi} \sin(\phi) \Delta V \Delta k_{\text{so}}  
\frac {8 k_{\text{F}}^2 L V_0}
{(4 k_{\text{F}}^2+V_0^2)^2}
\label{spinc}
\end{equation}
and noise 
\begin{equation}
\label{noise1} 
{S}_\sigma=\frac{|\omega|}{\pi}  \frac {2 k_{\text{F}}^2}
{(4 k_{\text{F}}^2+V_0^2)^2} \left( \Delta V^2 +\Delta k_{\text{so}}^2 L^2 
 V_0^2\right) ~.
\end{equation} 
For the particular pumping cycle chosen, and for vanishing 
temperature, the zero-frequency component of  
the spin-$\sigma$ current noise does not 
depend on the spin-index and on the phase $\phi$.
The spectrum of Eq.(\ref{noise1}) shows that the fluctuations 
introduced by the modulation of $\bar{V}$ and $k_{\text{so}}$ are 
uncorrelated. 
We can define a signal-to-noise ratio as 
$|I_{\text{spin}}|/ {S}_{\text{spin}}$ that in the present case  
reads 
\begin{equation} 
\label{sn}
\frac{|I_{\text{spin}}|}{S_{\text{spin}}}=
\frac{2}{\hbar} \frac{\left|
\sin(\phi) V_0^2  \Delta V \Delta k_{\text{so}} L \right|}
{ \Delta V^2 +\Delta k_{\text{so}}^2 L^2  V_0^2  }. 
\end{equation}
The signal-to-noise ratio Eq.~(\ref{sn}) reaches its maximum at fixed 
$\phi$ for 
$ \Delta V = \Delta k_{\text{so}} L  V_0$. 
   
In the simplest arrangement the spin-pumping effect can be detected 
if one of the two leads has been replaced by a half-metallic ferromagnet 
({\it i.e.} only majority spins are present). 
If its magnetization lies in the plane of the 
wire and makes an angle $\theta$ with the $y$-axis the spin state of the 
electrons in the ferromagnetic lead is $|F\rangle=\cos \frac{\theta}{2} 
|+\rangle+ i \sin \frac{\theta}{2} |-\rangle$ ($|\pm\rangle$ are the 
eigenstates of $\sigma_y$).
Furthermore, only to keep formulas compact, we assume that the 
Fermi velocity in the ferromagnetic lead is the same as in the 
rest of the system.
In this case the pumped charge and spin are given by 
\begin{eqnarray}
I^{\text{F}}_{\text{charge}}&=&\frac{I_{\text{spin}}}{2}\cos \theta  \\
I^{\text{F}}_{\text{spin}}&=&\frac{I_{\text{spin}}}{2}
\end{eqnarray}
($I_{\text{spin}}$ is the pumped spin current with normal leads).
There are several remarks that should be made: 
1) the spin current is independent of 
the magnetization  direction; 
2) the charge current is no more 
zero and it reaches its maximum when the magnetization is aligned with 
the spin-quantization axis in the wire; 
3) the charge current can be reversed changing $\theta$ into $\theta+\pi$. 
The dependence of the pumped charge on the magnetization direction can 
be exploited to verify that the pumping mechanism is taking place.

As a second possibility, 
we consider a magnetic 
field in the $y$-direction,
which introduces only a Zeeman term 
in the Hamiltonian $H_{B}=\frac{\hbar}{2} \Omega_{\text{B}} \sigma_y$. 
The effect of the Zeeman field is to modify the Fermi velocities for the 
two spin species. We can take this effect into account simply by 
replacing $k_{\text{F}}$ in Eq.~(\ref{dsigma}) with 
$k_{\text{F},\sigma}=  
k_{\text{F}}-\sigma \Delta k_{\text{F}}$, where $k_{\text{F}}$
is the Fermi wave-vector in the leads in the absence of 
magnetic field. 
Assuming that $|\Delta k_{\text{F}}|\ll k_{\text{F}}$ we can 
write $\Delta k_{\text{F}}=\frac{\Omega_{\text{B}}}{2}\frac{m}
{\hbar k_{\text{F}}}$ finding for $D_\sigma$ in the presence of the 
magnetic field 
(to lowest order in $\Delta k_{\text{F}}/k_{\text{F}}$)
\begin{equation}
\label{ab}
D_{\sigma}^{\text{B}}
(\bar{V},k_{\text{so}})= D_{\sigma}(\bar{V},k_{\text{so}})-
4 L \bar{V} \frac{4 k_{\text{F}}^2-\bar{V}^2 }
{(4 k_{\text{F}}^2+\bar{V}^2)^3} \frac{m \Omega_{\text{B}}}{\hbar}, 
\end{equation} 
where $D_{\sigma}$ is the expression in the absence of magnetic field 
given in Eq.~(\ref{dsigma}). The lowest order contribution in 
$\Delta k_{\text{F}}/k_{\text{F}}$ to the pumped spin current is zero, 
while the pumped charge current is 
 \begin{equation}
I^{\text{B}}_{\text{charge}}=-
\frac{\omega}{2\pi^2} \int d\bar{V} dk_{\text{so}} 8 L \bar{V} 
\frac{4 k_{\text{F}}^2-\bar{V}^2 }
{(4 k_{\text{F}}^2+\bar{V}^2)^3} \frac{m \Omega_{\text{B}}}{\hbar}. 
\end{equation}
The direction of charge flow can be reversed by changing the sign of 
$\Omega_B$, i.e. of the magnetic field. The detection of this 
effect would constitute an indirect evidence of spin pumping. 

\begin{figure}
\includegraphics[width=3.in]{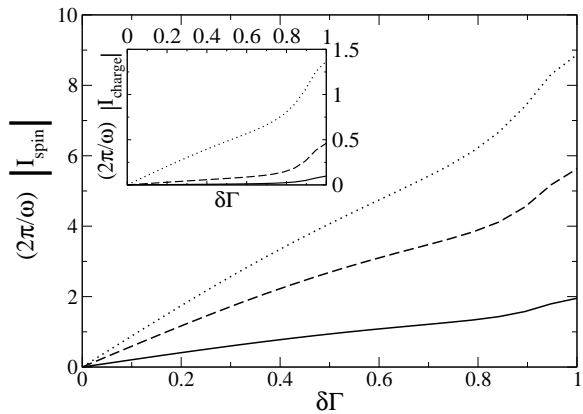}
\caption{Average spin and charge (in the inset) transmitted within a cycle
as a function of $\delta\Gamma$ for
$\gamma_{\text{so}}^{\text{max}}=0.042$ (full line),
$\gamma_{\text{so}}^{\text{max}}=0.125$ (dashed line)
and $\gamma_{\text{so}}^{\text{max}}=0.25$ (dotted line)
for fixed $\Gamma_{\text{0}}=0.5$.
For this choice of the parameters $I_{\text{spin}}$ and $I_{\text{charge}}$
have opposite sign.
The Fermi energy is set at $41/12$ (in units of $\gamma$).}
\label{plot}
\end{figure}

Until now we have studied an idealized model which allowed us 
to understand the physical phenomena that adiabatic spin pumping relies on. 
We now consider a more realistic model, which 
includes several modes, subband mixing induced by 
Eq.~(\ref{hmix}), and 
the effect of the time modulation of $\Delta_{\text{so}}$.
We numerically calculate the scattering matrix within the
tight-binding model, using a recursive Green's function technique~\cite{tech}.
The tight-binding version of the Rasha SO coupling can be written as~\cite{mireles}:
\begin{eqnarray}
H_{\text{so}}=-i\gamma_{\text{so}} \sum_{\sigma,\sigma'}\sum_{i,j} \left[
c^\dagger_{i+1,j,\sigma'} (\sigma_y)_{\sigma,\sigma'} ~c_{i,j,\sigma}-
\right.\\\nonumber \left.
-c^\dagger_{i,j+1,\sigma'}
(\sigma_x)_{\sigma,\sigma'} ~c_{i,j,\sigma}\right] + \text{h.~c}.~,
\end{eqnarray}
where $c^\dagger_{i,j,\sigma}$ is the creation operator of an electron
in site $(i,j)$ with spin $\sigma$ and $\gamma_{\text{so}}$ is the Rashba
nearest-neighbour coupling.
Note that $\gamma_{\text{so}}$ is related to the parameter $k_{\text{so}}$ through
the relation: $\gamma_{\text{so}}=(ak_{\text{so}})\gamma$, where $\gamma$ is the
tight-binding hopping potential and $a$ is the lattice constant.
In our simulations the wire is modeled as a 2D lattice with $W=3$ sites
in the transverse $y$-direction and $N=50$ sites in the longitudinal
$x$-direction.
The wire is then attached to the two leads, in which $\gamma_{\text{so}}=0$,
through a hopping potential $\Gamma_{\text{L}}$ on the left-hand-side and
$\Gamma_{\text{R}}$ on the right-hand-side (in the following we 
set $\Gamma_{\text{R}}=\gamma$).
The Fermi energy is chosen so that three bands are occupied
(from now on we express $\gamma_{\text{so}}$, $\Gamma_{\text{L}}$ and
$\Gamma_{\text{R}}$ in units of $\gamma$).
Adiabatic pumping is obtained by performing a square cycle in the
parameter space $(\gamma_{\text{so}},\Gamma_{\text{L}})$, with $\Gamma_{\text{L}}$
varying in the range $\Gamma_{\text{0}}\pm\delta\Gamma/2$
and $\gamma_{\text{so}}$ varying between zero and $\gamma_{\text{so}}^{\text{max}}$.
In Fig. \ref{plot} the average number of spins and charges transmitted in
a cycle are plotted as functions of
$\delta\Gamma$ for different values of $\gamma_{\text{so}}^{\text{max}}$ and
for fixed $\Gamma_{\text{0}}$.
As expected, both $I_{\text{spin}}$ and $I_{\text{charge}}$ are increasing
functions of $\delta\Gamma$.
It is remarkable that for $\gamma_{\text{so}}=0.042$ and
$\gamma_{\text{so}}=0.125$, corresponding to
typical values for Rashba splitting in semiconductor \cite{numbers},
$I_{\text{spin}}$
is about two orders of magnitude larger than $I_{\text{charge}}$
almost over the whole $\delta\Gamma$ range.
Note that even for $\gamma_{\text{so}}=0.25$, values which exceeds the
maximum reported Rashba coupling strength \cite{numbers}, $I_{\text{spin}}$
is still much larger than $I_{\text{charge}}$.
The pumped charge $I_{\text{charge}}$ remains much smaller compared to
$I_{\text{spin}}$ as long as we are in the weak Rashba coupling regime
\cite{mireles}, in which the inter-subband mixing due to Rashba coupling is
negligible (in our case $\gamma_{\text{so}}<0.38$).
Since in our simulations $\gamma_{\text{so}}\simeq 0.165$ corresponds
to the maximum reported value for $k_{\text{so}}$,
there is no need to go beyond the
weak Rashba regime, at least for narrow wires.
The inclusion of an additional constant on-site energy in the leads
(modeling a difference in the Fermi velocity between the leads and the wire)
does not introduce any new time dependence in the scattering matrix, and hence
it does not hinder the principle on which spin pumping is based on.
We also considered the presence of
disorder by adding to the tight-binding on-site energies in the Rashba region
 a random potential.
We find that averaging over disorder realizations yields a suppression
of the average $|I_{\text{spin}}|$, with respect to the clean case, but
keeping $|I_{\text{spin}}| \gg |I_{\text{charge}}|$.
In the quasi-ballistic regime \cite{note4}
adiabatic spin pumping still takes place with no qualitative difference.

We finally reanalyze the spin pumping from a different perspective.
To this end we start noticing that the Hamiltonian 
$H_{\text{1D}}$ [see Eq.~(\ref{h1d})] in the basis of eigenstates of 
$\sigma_y$ can be recast in the following form
$H_{\text{1D}}=\frac{1}{2m} (p_x-e \vec{A}_{\sigma,x}\cdot \hat{x})^2$,
where the spin-depended vector potential is given by 
$\vec{A}_{\sigma}=\frac{\hbar k_{\text{so}}}{e} \sigma\, \hat{x}$. 
As $\vec{\nabla} \times \vec{A}_\sigma=0$ this vector potential does not 
describe a magnetic field. But if  $k_{\text{so}}$ varies with 
time $\tau$ it describes a spin-dependent electric field 
$$\vec{E}_\sigma(\tau)=-\partial_\tau \vec{A}_\sigma(\tau)
=-\sigma \frac{\hbar }{e}
\partial_\tau k_{\text{so}}(\tau)\,\hat{x}~.$$
This electric field leads to a spin-dependent potential drop along 
the wire $V_\sigma=-\int_0^L \vec{E}_\sigma \cdot d\vec{x} = 
\sigma \frac{\hbar} {e} \partial_\tau k_{\text{so}} L$. 
Thus, we can consider the wire without SO coupling but with 
spin-$\sigma$ electrons subject 
to the time-dependent potential $V_\sigma$.
An additional  time-dependent barrier $V_{\text{bar}}$ (as 
it was the case for the pumping cycle that led to Eq.~(\ref{dsigma}))
leads to  rectification of the the oscillating potential $V_\sigma$. 
Provided that the voltage $V_\sigma$ is small enough so that 
linear transport theory applies, and that it changes on 
a time scale much larger than the time an electron needs to 
go through the scattering region the average spin-$\sigma$ current reads
\cite{butti}
\begin{equation}
\label{recti}
I_\sigma=\frac{\omega}{2 \pi}\frac{e}{h}\int_0^{\frac{2 \pi}{\omega}} 
\left| t(\tau)\right|^2 V_\sigma(\tau) d\tau, 
\end{equation}   
where $t$ is the transmission coefficient for $k_{\text{so}}=0$.
From
Eq.~(\ref{recti}) all the results 
obtained so far can be found. 
In this framework spin pumping appears as a rectification effect \cite{note3}.
Furthermore, the gauge transformation to spin-dependent electric fields 
shows that the pumping mechanism survives when no phase coherence 
is present (although the description that relies on the scattering matrix 
ceases to be valid). The spin current will be limited by the spin-relaxation 
rate (which depends on temperature).    
 
In conclusion we have shown that spin  currents can be produced through 
adiabatic pumping with no use of ferromagnets or magnetic fields.
Only electrical gating and the tunability of the Rashba coupling
are exploited.
This effect is roubust also when several propagating modes are present,
even in the case of non-negligible subband mixing and disorder.
In addition, two possible different ways to detect spin pumping have been
discussed and the zero-frequency noise spectrum has been calculated.

We gratefully acknowledge helpful discussions with M.~B\"uttiker,  
L.S. Levitov, and Yu. Makhlin.
This work was supported by the EU (IST-SQUBIT), RTN2-2001-00440,
HPRN-CT-2002-00144 and by the 
German Science Foundation (DFG) through the Center for Functional 
Nanostructures in Karlsruhe.


\begin{thebibliography}{24}
\bibitem{wolf}
        S.~A.~Wolf, D.~D.~Awschalom, R.~A.~Buhrman, J.~M.~Daughton,
	S.~von~Moln\'ar, M.~L.~Roukes, A.~Y.~Chtchelkanova, and
	D.~M.~Treger, Science {\bf 294}, 1488 (2001). 
\bibitem{dassarma}
        S. Das Sarma, S.~J.~Fabian, X.~Hu, and I.~Zutic, Sol. Stat. Comm. {\bf 119}, 207 (2001).
\bibitem{ganichev} S.~D. Ganichev, E.~L.~Ivanchenko, S.~N.~Danilov, J.~Eroms,
	W.~Wegscheider, D.~Weiss, and W.~Prettl,  Phys. Rev. Lett. {\bf 86},
        4358  (2001).
\bibitem{mucciolo}
        E.~R. Mucciolo, C. Chamon, C.M. Marcus, Phys. Rev. Lett. 
        {\bf 89}, 146802 (2002).
\bibitem{brataas}
        A. Brataas, Y.~Tserkovnyak, G.~E.~W.~Bauer, and B.~I.~Halperin,
	Phys. Rev. B {\bf 66}, 060404(R) (2002).
\bibitem{brouwer} 
        P.~W.~Brouwer, Phys. Rev. B {\bf 58}, R10135 (1998). 
\bibitem{zhou} 
        F.~Zhou, B.~Spivak, and B.~Altshuler, Phys. Rev. Lett. 
        {\bf 82}, 608 (1999). 
\bibitem{switkes} 
        M.~Switkes, C.~M.~Marcus, K.~Campman, and A.~C.~Gossard, Science {\bf 283}, 1905 (1999). 
\bibitem{makhlin}
        Yu. Makhlin and A.~D.~Mirlin, Phys. Rev. Lett. {\bf 87}, 
        276803 (2001).
\bibitem{levitov} L.~S.~Levitov, cond-mat/0103617 (2001)
\bibitem{levinson} O.~Entin-Wohlman, A.~Aharony, and  
        Y. Levinson, Phys. Rev. B {\bf 65}, 195411 (2002).
\bibitem{aleiner}
        I.~L. Aleiner, B.L. Altshuler and A. Kamenev,
        Phys. Rev. B {\bf 62}, 10373 (2000).
\bibitem{moskalets} M.~Moskalets, and M.~B\"uttiker, Phys. Rev. B {\bf 66}, 
035306 (2002).
\bibitem{butti} M.~Moskalets, and M.~B\"uttiker, Phys. Rev. B {\bf 64}, 
201305 (2001).
\bibitem{rashba} E.~I.~Rashba, Fiz. Tverd. Tela (Leningrad) {\bf 2}, 1224 
(1960), [Sov. Phys. Solid State {\bf 2}, 1109 (1960)].
\bibitem{lommer} G.~Lommer, F.~Malcher, and U. R\"ossler, Phys. Rev. Lett. 
{\bf 60}, 728 (1988)
\bibitem{note0} Several devices have been proposed that make use of the 
Rashba spin-orbit coupling and of its tunability to obtain interesting
spin-dependent effects:  
S. Datta and B. Das, Appl. Phys. Lett. {\bf 56},665 (1990);
E.~A.~de Andrada e Silva and G.~C.~La Rocca, Phys. Rev. B {\bf 59}, R15583 
(1999); A.~A.~ Kiselev and K.~W.~Kim, Appl. Phys. Lett. {\bf 78},  
775, (2001); T.~Koga, J.~Nitta, H.~Takayanagi, and S.~Datta, Phys. Rev. Lett. {\bf 88}, 126601 (2001); 
M.~Governale, D.~Boese, U.~Z\"ulicke, and C.~Schroll, Phys. Rev. B {\bf 65}, 140403(R) (2002);
R.~Ionicioiu and I.~D'Amico,  Phys. Rev. B {\bf 67}, 041307(R) (2003);
J.~C.~Egues, G.~Burkard, and  D.~Loss, Phys. Rev. Lett. {\bf 89}, 
176401 (2002); M. Governale, Phys. Rev. Lett., {\bf 89}, 206802 (2002);
L.~S.~Levitov and E.~I.~Rashba, Phys. Rev. B {\bf 67}, 115324 (2003). 
\bibitem{nitta} J.~Nitta, T.~Akazaki, H.~Takayanagi, and T.~Enoki, 
Phys. Rev. Lett. {\bf 78}, 1335 (1997). 
\bibitem{schaepers} T.~Sch\"apers, G.~Engels, J.~Lange, Th.~Klocke,
	M.~Hollfelder, and H.~L\"uth, J. Appl. Phys. {\bf 83}, 
4324 (1998).  
\bibitem{grundler} D.~Grundler, Phys. Rev. Lett. {\bf 84}, 6074 (2000). 
\bibitem{mireles} F.~Mireles, and G.~Kirczenow, Phys. Rev. B {\bf 64}, 
024426 (2001). 
\bibitem{rashbawire} M.~Governale, and U.~Z\"ulicke, 
Phys. Rev. B {\bf 66}, 073311 (2002).
\bibitem{note1} If $k_{\text{so}}$ depends on time also 
$\Delta_{\text{so}}$ does. The only effect of the time variation of 
 $\Delta_{\text{so}}$ is to induce some small charge pumping, but it 
does not affect spin-pumping.  
\bibitem{note2}
The reflection and transmission coefficients for the present model, 
which are needed to compute $D_\sigma$, read     
$t_\sigma=2 k_{\text{F}}e^{i \sigma k_{\text{so}} L}/(2k_{\text{F}}+i
 \bar{V})$, 
$r'_\sigma=-i\, {\bar{V}}/ (2k_{\text{F}}+i \bar{V})$. 
\bibitem{tech} S. Sanvito, C.~J. Lambert, J.~H. Jefferson,
        and A.~M. Bratkovsky, Phys. Rev. B {\bf 59}, 11936 (1999).
\bibitem{numbers} Typically for InGaAs:
$m=0.042m_{\text{0}}$, where $m_0$ is the electron bare mass, $a=10$ nm
and $k_{\text{so}}$ ranges from 0.33 to 1.65 $\times 10^7$ m$^{-1}$,
corresponding to $\gamma_{\text{so}}=0.033$ and $\gamma_{\text{so}}=0.165$.
\bibitem{note4} In the simulations (not reported here) we take the 
on-site energy randomly distributed in the range $[-U/2, U/2]$ 
with $U$ as large as $1/10~E_\text{F}$. 
\bibitem{note3} Further predictions can be made using Eq. (\ref{recti}).
If in a time $T$, the SO coupling strength 
goes from zero to a value $\Delta k_{\text{so}}$ and that for 
$\tau > T$ it stays fixed at this value,
during the interval $T$ a net spin 
current flows (but no charge current), and the total spin transferred  
is simply given by 
$\frac{2e}{h} |t|^2 \int_0^T V_+(\tau) d\tau$. If one of the leads is 
replaced by a ferromagnet or a transverse in-plane magnetic field is present, 
this particular variation of $ k_{\text{so}}$ 
will result in a pulse of charge current of duration $T$.   
\end{thebibliography}
\end{document}